\documentclass[aps,prl,twocolumn,superscriptaddress,showpacs]{revtex4-1}
\usepackage{graphicx}
\usepackage{float}
\usepackage{gensymb}
\usepackage{color}
\usepackage{dcolumn}   
\usepackage{bm}        
\usepackage{amssymb}   
\usepackage{amsmath}
\usepackage{textcomp}
\usepackage{bbm}
\usepackage{hyperref}
\usepackage{graphicx}
\usepackage{array}
\usepackage{braket}
\usepackage{multirow}
\usepackage{natbib}
\usepackage{verbatim}

\begin{document}


\title{Crystal field excitations from $\mathrm{\mathbf{Yb^{3+}}}$ ions at defective sites in highly stuffed $\mathrm{\mathbf{Yb_2Ti_2O_7}}$}

\author{G. Sala}
\email{salag@ornl.gov}
\affiliation{Neutron Scattering Division, Oak Ridge National Laboratory, Oak Ridge, Tennessee 37831, USA}
\affiliation{Department of Physics and Astronomy, McMaster University, Hamilton, ON L8S 4M1 Canada}

\author{D. D. Maharaj}
\affiliation{Department of Physics and Astronomy, McMaster University, Hamilton, ON L8S 4M1 Canada}

\author{M. B. Stone}
\affiliation{Neutron Scattering Division, Oak Ridge National Laboratory, Oak Ridge, Tennessee 37831, USA}

\author{H. A. Dabkowska}
\affiliation{Brockhouse Institute for Materials Research, McMaster University, Hamilton, ON L8S 4M1 Canada}

\author{B. D. Gaulin}
\email{bruce.gaulin@gmail.com}
\affiliation{Department of Physics and Astronomy, McMaster University, Hamilton, ON L8S 4M1 Canada}
\affiliation{Brockhouse Institute for Materials Research, McMaster University, Hamilton, ON L8S 4M1 Canada}
\affiliation{Canadian Institute for Advanced Research, 661 University Avenue, Toronto, ON,  M5G 1M1 Canada}





\date{\today}

\begin{abstract}
The pyrochlore magnet Yb$_2$Ti$_2$O$_7$ has been proposed as a quantum spin ice candidate, a spin liquid state expected to display emergent quantum electrodynamics with gauge photons among its elementary excitations.  However, Yb$_2$Ti$_2$O$_7$'s ground state is known to be very sensitive to its precise stoichiometry.  Powder samples, produced by solid state synthesis at relatively low temperatures, tend to be stoichiometric, while single crystals grown from the melt tend to display weak ``stuffing" wherein $\mathrm{\sim 2\%}$ of the $\mathrm{Yb^{3+}}$, normally at the $A$ site of the $A_2B_2O_7$ pyrochlore structure, reside as well at the $B$ site.  In such samples $\mathrm{Yb^{3+}}$ ions should exist in defective environments at low levels, and be subjected to crystalline electric fields (CEFs) very different from those at the stoichiometric $A$ sites.  New neutron scattering measurements of $\mathrm{Yb^{3+}}$ in four compositions of Yb$_{2+x}$Ti$_{2-x}$O$_{7-y}$, show the spectroscopic signatures for these defective $\mathrm{Yb^{3+}}$ ions and explicitly demonstrate that the spin anisotropy of the $\mathrm{Yb^{3+}}$ moment changes from XY-like for stoichiometric $\mathrm{Yb^{3+}}$, to Ising-like for ``stuffed" B-site $\mathrm{Yb^{3+}}$, or for A-site $\mathrm{Yb^{3+}}$ in the presence of an oxygen vacancy.
\end{abstract}




\maketitle


Exotic magnetic ground states of cubic pyrochlore magnets, with composition $A_2B_2O_7$, are of great topical interest, as the pyrochlore lattice is one of the canonical architectures supporting geometrical frustration in three dimensions \cite{Lacroix, Greedan}.  Magnetism can reside at either the A$^{3+}$ site or the B$^{4+}$ site, and the magnetic moments' anisotropy and the interactions between the moments conspire to give rise to rich ground state selection.  Among the states and materials that have been of recent interest have been the classical spin ice states in Dy and Ho titanate pyrochlores \cite{Harris1997,Ramirez,DenHertog2000,Bramwell2001,Castelnovo2008}, spin liquid and spin glass states in molybdate pyrochlores \cite{Lucy}, and spin fragmentation in Nd based zirconate pyrochlores \cite{Petit}. The possibility that a quantum analogue of the spin ice ground state, i.e. quantum spin ice (QSI), may exist in certain low moment pyrochlore magnets, including Yb$_2$Ti$_2$O$_7$ and Pr$_2$Zr$_2$O$_7$, has generated much excitement \cite{Thompson2,Thompson,RossPRX,SavaryPRL,Gingras,Benton,Hayre,SavaryPRB,DOrtenzio,Kimura,Gingras2,LiDong,Robert,LiDong2,Hamachi}.

\begin{figure}
\includegraphics[width=0.9\columnwidth]{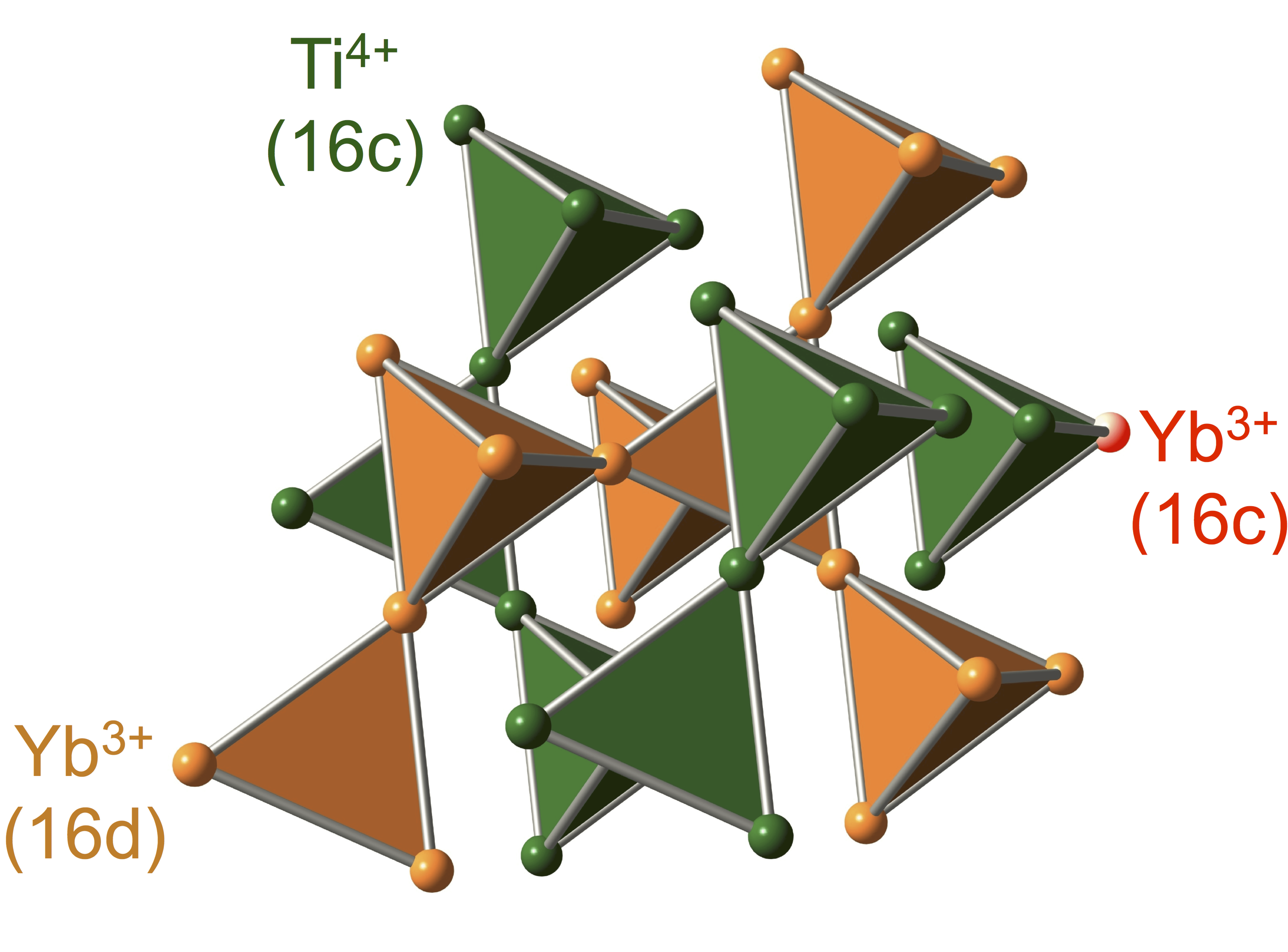}
\caption{\label{fig: 1}(color online) The pyrochlore lattice, displayed by $A_2B_2O_7$ compounds, belongs to the $Fd\bar{3}m$ space group and consists of two interpenetrating networks of corner-sharing tetrahedra.  In stoichiometric Yb$_2$Ti$_2$O$_7$, the $A$ sublattice is occupied by rare-earth magnetic $\mathrm{Yb^{3+}}$ ions (orange spheres) and the $B$ sublattice is occupied by  nonmagnetic $\rm Ti^{4+}$ site (green spheres). In stuffed Yb$_{2+x}$Ti$_{2-x}$O$_{7-y}$, a small fraction of $\mathrm{Yb^{3+}}$ ions (in red) also occupy the $B$ sites and they experience a different crystalline electric field due to the different local environment of surrounding ligands at the B site, compared with the A site.
}
\end{figure}

At low temperatures Yb$_2$Ti$_2$O$_7$ displays two magnetic heat capacity anomalies: a broad one near 2 K and a sharp anomaly signifying a thermodynamic phase transition near $T_C$ = 0.26 K \cite{Blote, deReotier, Yaouanc, RossStuffing, ARCMP}. Below $T_C$, the ordered structure is thought to be a splayed ferromagnet with moments pointing close to the (100) directions \cite{Yasui,Chang2012,GaudetGapless,Yaouanc3,ScheieLobed}.  However, surprising sample variability has been reported in this phase transition, with some studies not seeing direct evidence for the ferromagnetic ordered state \cite{Ross2009,RossDimensional,Hodges2002,Yaouanc2,Gardner,DOrtenzio,Bhattacharjee,Bonville}.  Using the sharp anomaly in C$_P$ as the figure-of-merit for the phase transition, interesting systematics have been observed \cite{RossStuffing,ARCMP,Arpino,Ali}. Powder samples grown by solid state synthesis at relatively low temperatures show a sharp C$_P$ anomaly and a high $T_C$. usually $\sim$ 0.26 K \cite{RossStuffing,ARCMP,Arpino,Antonio}; however most single crystal studies display broader thermodynamic anomalies at much lower temperatures, often with $T_C$s around and below 0.2 K \cite{Coldea,Lhotel,Yasui,Chang2012}.

Crystallographic studies of the powder and single crystal samples have revealed that the powder samples are stoichiometric Yb$_2$Ti$_2$O$_7$, while the single crystals are ``lightly stuffed", and characterized by the composition Yb$_{2+x}$Ti$_{2-x}$O$_{7-y}$, with $x$ $\sim$ 0.04 \cite{RossStuffing}. That is, a small excess of $\mathrm{Yb^{3+}}$ ions, nominally at the crystallographic 16\textit{d} or \textit{A} site, are ``stuffed" onto the 16\textit{c} or \textit{B} site where nonmagnetic $\mathrm{Ti^{4+}}$ ions are located in pure Yb$_2$Ti$_2$O$_7$ as schematically indicated in Fig.~\ref{fig: 1}. Light stuffing is also known to occur in other titanate pyrochlores \cite{Baroudi}.

It is remarkable that such a small change in stoichiometry could so strongly effect the ground state selection of a simple ordered state in a three dimensional magnetic insulator.  Related phenomena has also recently been observed in the effect of hydrostatic pressure on stoichiometric Yb$_2$Ti$_2$O$_7$ samples, where ambient pressure conditions show no sign of a $\mu$SR signal for the transition, but a minimal 1 kbar (and above) applied pressure results in a clear signal for a transition near $T_C$ $\sim$ 0.26 K \cite{Edwin}.  

With weak ``stuffing"  able to suppress this phase transition by as much as $\mathrm{\sim 25\%}$ \cite{RossStuffing,ARCMP,Arpino,Yasui,Chang2012}, it is important to understand precisely what is at play in its ground state selection.  One thing that is clear is that most single crystals of Yb$_2$Ti$_2$O$_7$ likely have Yb$^{3+}$ ions occupying not only the stoichiometric A-sites, but also B-sites.  They also possess A-sites with missing oxygen neighbours. The Yb$^{3+}$ ions in defective environments are expected to experience very different crystal field effects than those at stoichiometric A-sites \cite{Gaudet}. As these effects determine the spin anisotropy and size of the Yb$^{3+}$ moment, it is possible that the defective Yb$^{3+}$ moments and their anisotropy are very different from those displayed by stoichiometric Yb$^{3+}$ - indeed a prediction from point charge calculations of the crystal field effects on Yb$^{3+}$ have suggested that this is the case \cite{Gaudet}.

The eigenvalues and eigenfunctions associated with crystal field states can be determined using inelastic neutron spectroscopy, and these have been determined for stoichiometric Yb$_2$Ti$_2$O$_7$ and several other rare-earth based pyrochlore magnets \cite{Gaudet,Rosenkranz,Bertin}.  However, the equivalent measurements on Yb$^{3+}$ in defective environments in Yb$_{2+x}$Ti$_{2-x}$O$_{7-y}$ are much more difficult, as the environments occur at low density in these materials.  Additionally, as we will see, the eigenvalues associated with the defective environments tend to extend to much higher energies.

Powder samples of Yb$_{2+x}$Ti$_{2-x}$O$_{7-y}$ with x = 0.11 and 0.18 were prepared at McMaster University and characterized using the POWGEN neutron powder diffractometer \cite{POWGEN} at the Spallation Neutron Source of Oak Ridge National Laboratory. Our best refinement of this powder diffraction data gives x = 0.106(4) and 0.176(8) for the highly stuffed samples with oxygen vacancies preferentially located at the O(1) site of the pyrochlore lattice, as discussed in the Supplemental Material (SM).

Inelastic neutron scattering measurements were performed on these two highly stuffed powder samples. The resulting samples were $\approx 4$ g of powder for each of the x = 0.11
and 0.18 samples.  We studied their CEF excitations using the direct geometry time-of-flight spectrometer SEQUOIA~\cite{Sequoia} at ORNL and compared these results with earlier measurements performed \cite{Gaudet} on the stoichiometric (x = 0) and lightly stuffed (x = 0.05) samples.  The powder samples were loaded into aluminium flat plates and were sealed under He atmosphere in a glove box. An empty, aluminium flat plate with the same dimensions was prepared in a similar manner and employed for background measurements. Measurements have been performed at T = 5 K and 300 K, with incident energies of $\mathrm{E_i}$ = 150, 250 and 500~meV. The corresponding chopper settings selected at these energies were: $\mathrm{T_0 = 150~Hz}$ and $\mathrm{FC_2 = 600~Hz}$, $\mathrm{T_0 = 120~Hz}$ and $\mathrm{FC_2 = 600~Hz}$ and  $\mathrm{T_0 = 150~Hz}$ and $\mathrm{FC_2 = 600~Hz}$ respectively. The data were reduced with Mantid~\cite{mantid} and analyzed using DAVE~\cite{dave} software, while we employed custom software to refine the CEF spectrum of the powder samples, as described above and in the SM.

The CEFs originate primarily from the ``cage" of O$^{2-}$ ions surrounding the cations, lifting the (2J+1)-fold degeneracy of the J = 7/2 ground state manifold appropriate to $\mathrm{Yb^{3+}}$.  Typical time-of-flight inelastic neutron scattering data from four powder samples of Yb$_{2+x}$Ti$_{2-x}$O$_{7-y}$, with x = 0, 0.05, 0.11, and 0.18 are shown at T = 5 K in Figs. 3, and for x = 0.11 and 0.18 in Fig. 4.  This data has been analysed from a starting point,  known as the point charge approximation ~\cite{Hutchings, Freeman1962, Malkin2004, Walter1984}, where initial CEF transitions from the ground state are calculated based on the local symmetry of the $A$ and $B$ site $\mathrm{Yb^{3+}}$ ions, taken from crystallographic measurements. The energies and intensities of the CEF transitions are subsequently refined to agree with the experimental data. The case for $\mathrm{Yb^{3+}}$ is relatively straightforward as its 13 4\textit{f} electrons give a Hund's rule ground state of J = 7/2, so it is a Kramers' ion with at most 3 CEF transitions from the ground state.

\begin{figure}
\includegraphics[width=0.9\columnwidth]{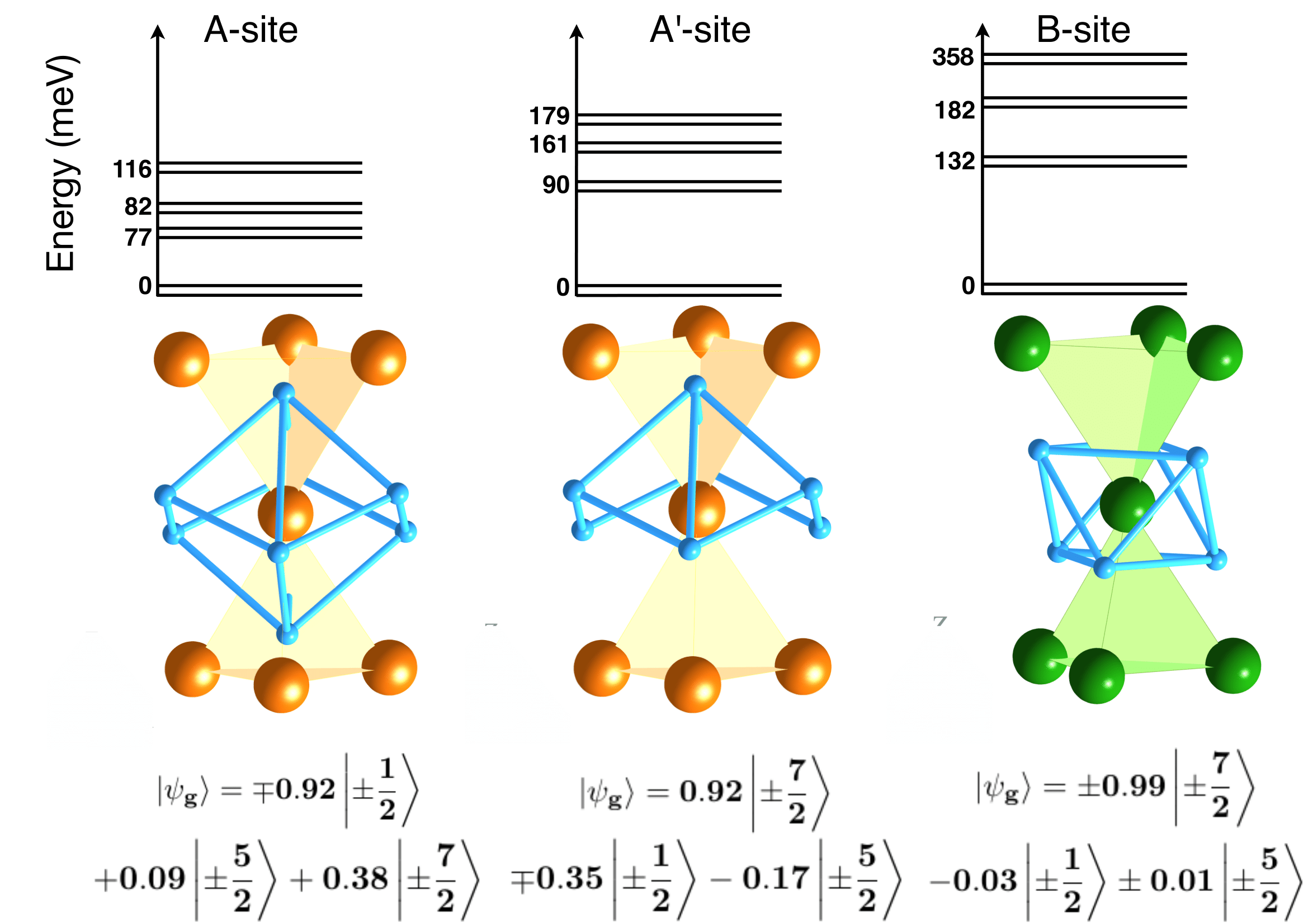}
\caption{\label{fig: 2}(color online)\textbf{(left)} At the stoichiometric $A$ sites, the ligands are located on the vertex of a scalenohedron, a cube distorted along the diagonal. The blue spheres represent the six O(2) ions and two O(1) ions  around the $A$ site $\mathrm{Yb^{3+}}$.  The O(1) sites are located along the axis connecting the centres of two tetrahedra.  \textbf{(center)} The $A^\prime$ sites correspond to A-sites with one O(1) vacancy that breaks the symmetry of the scalenohedron. \textbf{(right)} At the $B$ sites, the environment is a trigonal anti-prism made of six O(2) ions; green spheres represent the position of Ti$^{4+}$ ions, or the position of the stuffed $\mathrm{Yb^{3+}}$ ions. The top panels show the corresponding energy eigenvalues associated with each environment. Note that the energy scale is approximate and serves only to guide the eye for comparison of the CEF excitation energies.  The bottom panels give the three largest contributions to the ground state energy eigenfunctions associated with each environment.
}
\end{figure}

We considered 3 local Yb$^{3+}$ environments shown in Fig.~\ref{fig: 2}. These are Yb$^{3+}$ in an \textit{A} site environment with a full complement of 8 neighbouring O$^{2-}$ ions; in an \textit{A} site 
environment with one O$^{2-}$ vacancy (referred to as an $A^\prime$ site); and a Yb$^{3+}$ ion in a \textit{B} site environment with a full complement of 6 neighbouring O$^{2-}$ ions.  The \textit{A} site O$^{2-}$ environment 
consists of a cube distorted along the local [111] directions. Six O(2) ions are located on a plane perpendicular to this direction and a three-fold rotation axis. Two additional O(1) ions are located along the local [111] axis. In other titanate pyrochlores, the O(1) sites are known to have a higher probability of hosting vacancies than the O(2) sites \cite{Sala2014}, a result which we confirm here for Yb$_{2+x}$Ti$_{2-x}$O$_{7-y}$ using powder neutron diffraction, as shown in the SM. By contrast, the environment at the \textit{B} site is a trigonal anti-prism made of six O(2) oxygen ions.  Additional local Yb$^{3+}$ environments, such as an \textit{A} site Yb$^{3+}$ with two vacancies, were assumed to be unlikely at the stuffing levels considered here.

Our CEF calculation followed Prather's convention~\cite{Prather} and employed the Stevens' formalism~\cite{Stevens}.
The resulting CEF Hamiltonian for $\rm Yb^{3+}$ at all three sites can be written as:
%
\begin{align}
\mathcal{H}_{CEF} = B^0_2\hat{O}^0_2 + B^0_4\hat{O}^0_4 +  B^3_4\hat{O}^3_4 + B^0_6\hat{O}^0_6 + B^3_6\hat{O}^3_6 + B^6_6\hat{O}^6_6
  & \label{eq: HCEF}
\end{align}

Here $\hat{O}^m_n$ are the Stevens' operators and $B^m_n$ the CEF parameters used to approximate the Coulomb potential generated by the ligands. 

The unpolarised neutron partial differential magnetic cross-section can be written within the dipole approximation as~\cite{Squires}:
\begin{eqnarray}
\frac{d^2\sigma}{d\Omega dE'} = C\frac{k_f}{k_i}F^2(|Q|)S(|Q|,\omega)
\end{eqnarray} 
where $\Omega$ is the scattered solid angle, $\frac{k_f}{k_i}$ the ratio of the scattered and incident momentum of the neutron, C is a constant and $F(|Q|)$ is the magnetic form factor of the magnetic Yb$^{3+}$ ion. 
The scattering function $S(|Q|,\hbar \omega)$ gives the relative scattered intensity due to transitions between different CEF levels. At constant temperature and wave vector $|Q|$, we have:
%
\begin{align}
S(|Q|,\hbar\omega) = \sum_{i,i'}\frac{(\sum_{\alpha}  |\langle i {| J_{\alpha} | i'\rangle |}^2) \mathrm{e}^{-\beta E_{i}} }{\sum_j \mathrm{e}^{-\beta E_{j}}} F(\Delta E + \hbar \omega)
  & \label{eq: SQw}
\end{align}

where $\alpha = x,y,z$ and $F(\Delta E + \hbar \omega) = F(E_{i} - E_{i'} + \hbar \omega)$ is a Lorentzian function which ensures energy conservation as the neutron induces transitions between the CEF 
levels $i \rightarrow i'$, that possess a finite energy width or inverse lifetime.

\begin{figure}
\includegraphics[width=0.9\columnwidth]{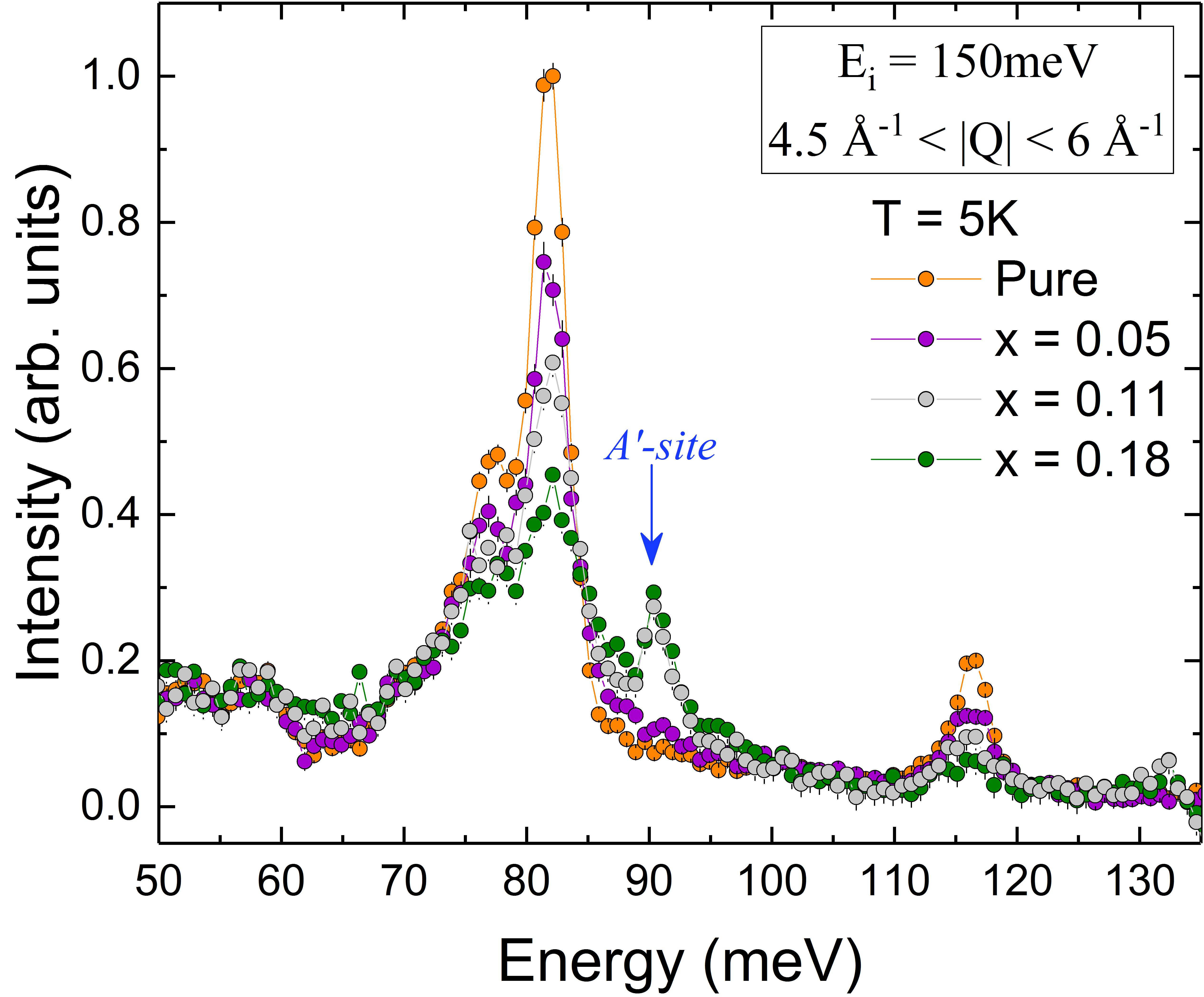}
\caption{\label{fig: 3}(color online) The measured neutron scattering intensity obtained from $\mathrm{E_i = 150~meV}$ data sets for four samples of Yb$_{2+x}$Ti$_{2-x}$O$_{7-y}$ is shown.  An empty can data set has been subtracted from all data. The difference between the CEF intensities of the four samples is evident. In the stoichiometric compound (orange points) there are only three visible levels at $\sim$ $76$, $81$ and $116$ meV; their intensities decrease as the system is ``stuffed" (as $x$ increases) and a new CEF level at $\sim$ $90$ meV appears. We estimate the ``stuffing level", $x$, both crystallographically and by comparing the intensity of the $90$ meV CEF level with the results of a MC simulation (see SM for details). The intensities have been scaled in proportion to the actual sample masses.
}
\end{figure}

Figure~\ref{fig: 3} shows a comparison of the data from the four powder Yb$_{2+x}$Ti$_{2-x}$O$_{7-y}$ samples, using incident neutrons with $\mathrm{E_i = 150 meV}$.  The intensity scale has 
been normalized to sample mass.  The stoichiometric, x = 0, and lightly stuffed, x = 0.05, powder samples show only the 3 \textit{A} site CEF transitions at $\sim$ 76, 81, and 116 meV as previously reported \cite{Gaudet}.  
As a function of increasing stuffing, x, we clearly observe the growth of a new CEF at $\sim$ 91 meV, which we will attribute to $A^\prime$ site Yb$^{3+}$.  A Monte Carlo calculation, shown in the SM, of the prevalence 
of \textit{A} to $A^\prime$ site Yb$^{3+}$ ions as a function of x, and with y in Yb$_{2+x}$Ti$_{2-x}$O$_{7-y}$ set to ensure charge neutrality, shows that the normalized intensity of this 91 meV CEF scales in proportion to x.  

\begin{figure}
\includegraphics[width=0.9\columnwidth]{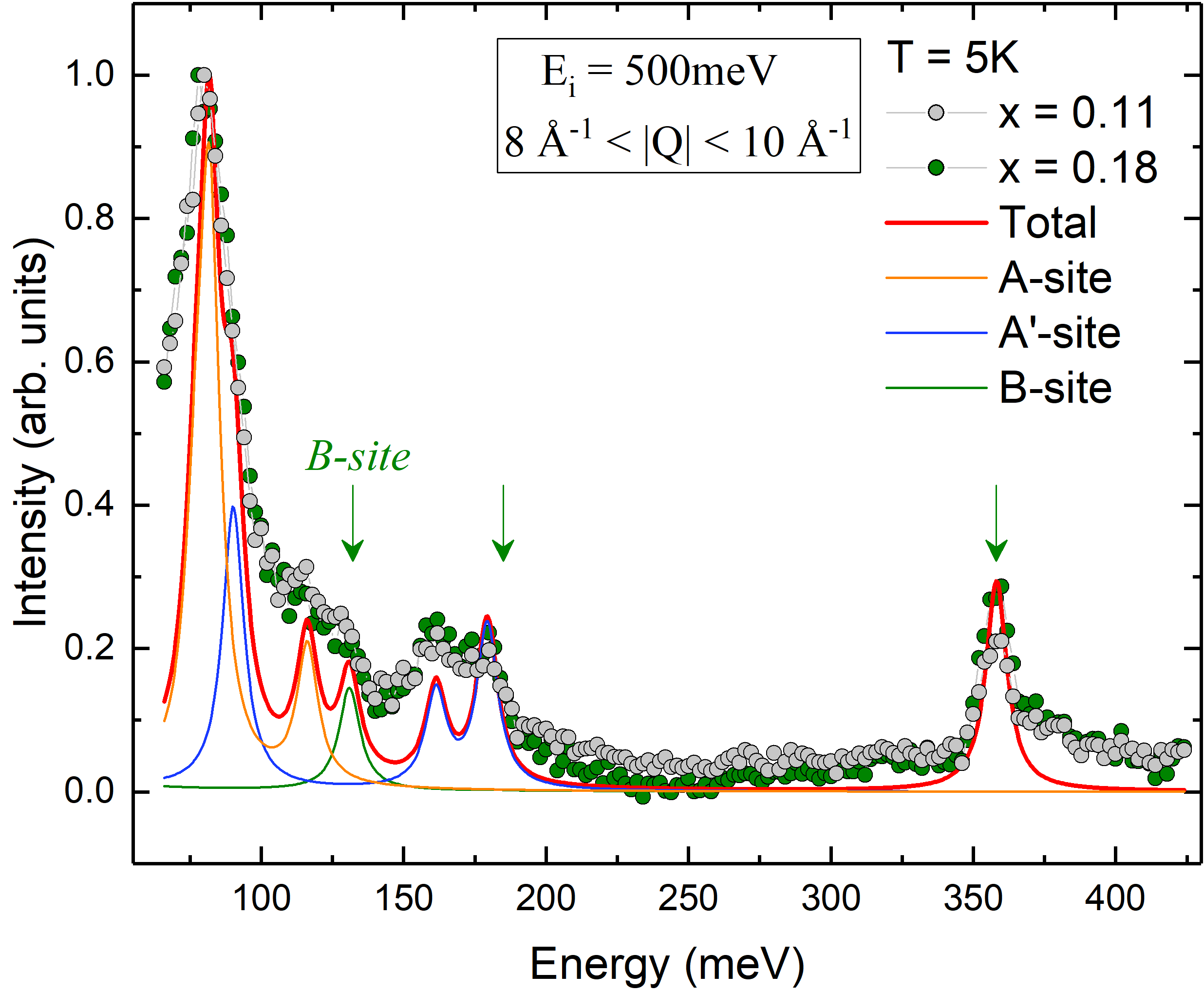}
\caption{\label{fig: 4}(color online) The measured neutron scattering intensity obtained from $\mathrm{E_i = 500~meV}$ data sets for two samples of Yb$_{2+x}$Ti$_{2-x}$O$_{7-y}$ are shown. An empty sample can subtraction has been empoyed. The calculated spectrum (red line) shows good agreement with the experimental data (gray and green dots for the x = 0.11 and x = 0.18 samples, respectively). The 3 different contributions to the total spectrum are highlighted in orange for the stoichiometric $A$ sites, blue for the oxygen deficient $A^\prime$ sites and green for stuffed $\mathrm{Yb^{3+}}$ ions at the $B$ sites. The total calculated intensity from all three sites is shown in red.
}
\end{figure}

The CEF spectrum at energies above 100 meV is shown in Fig.~\ref{fig: 4}, for the x = 0.11 and 0.18 powder samples, as measured with $\mathrm{E_i = 500 meV}$ neutrons.  One observes clear excitations above the 116 meV CEF 
excitation associated with the stoichiometric \textit{A} site's most energetic CEF level.  Of particular note is the well isolated CEF excitation at 358 meV which we associate with Yb$^{3+}$ at the \textit{B} site, and whose intensity 
scales between the x = 0.11 and 0.18 powder samples in proportion to x.  The stoichiometric (x = 0) and lightly stuffed (x = 0.05) powder samples were measured at high energies with E$_i$ = 700 meV neutrons, and the 358 meV 
CEF excitation is not visible for either.

The energies and relative intensities of all the CEF excitations measured below $\sim$ 400 meV were fit as described above, assuming the CEF parameters previously established for the stoichiometric x = 0 
sample \cite{Gaudet}. The results are shown as the solid lines in Fig.~\ref{fig: 4}, with the new CEF parameters and energies for the $A^\prime$ site and \textit{B} site tabulated along with those of the \textit{A} site Yb$^{3+}$ in 
the SM.  The description of all CEF levels below $\sim$ 400 meV is very good, and the resulting CEF energy eigenvalues are shown for the \textit{A}, $A^\prime$ and \textit{B} site Yb$^{3+}$ in the 
top panel of Fig.~\ref{fig: 2}. The bandwidth of the CEF excitations is much larger for Yb$^{3+}$ in the defective environments, with the defective \textit{B} site environment giving the largest bandwidth, consistent with this Yb$^{3+}$ ion experiencing the largest electric fields and their gradients.

The determination of the CEF parameters allows a determination of the g-tensor characterising the anisotropy, as well as the moment size associated with the ground state doublet of  Yb$^{3+}$ at the \textit{A}, $A^\prime$ 
and \textit{B} sites.  The resulting eigenfunctions within the Yb$^{3+}$ ground state doublets are shown in the bottom panel of Fig.~\ref{fig: 2}. The corresponding anisotropic g-tensor values are $\mathrm{g_{\perp} = 3.69 \pm 0.15}$, 
$\mathrm{g_z = 1.92 \pm 0.20}$ for Yb$^{3+}$ at the \textit{A} site; $\mathrm{g_{\perp} = 1.5 \pm 0.2}$, $\mathrm{g_z = 6.8 \pm 0.7}$ for Yb$^{3+}$ at the $A^\prime$ site; and $\mathrm{g{\perp} = 0.07 \pm 0.03}$, 
$\mathrm{g_z = 8.0 \pm 0.8}$ for Yb$^{3+}$ at the \textit{B} site.  The \textit{A} site Yb$^{3+}$ moment was previously known to display XY anisotropy \cite{Gaudet}. These results show both the $A^\prime$ site and \textit{B} 
site Yb$^{3+}$ moments to possess Ising-like anisotropies, with the \textit{B} site Yb$^{3+}$ Ising anisotropy being stronger than that associated with the $A^\prime$ site.  Such a change in anisotropy between stoichiometric 
and defective Yb$^{3+}$ sites was predicted on the basis of point charge calculations, but has now been directly verified with these measurements \cite{Gaudet}.  The ground state moment associated with the \textit{A}, $A^\prime$ 
and \textit{B} sites are found to be $\mu = 2.07\mu_B$, $\mu = 3.5\mu_B$ and $\mu = 4.0\mu_B$, respectively\cite{Gaudet}.  While dipolar interactions are expected to be relatively weak in Yb$_2$Ti$_2$O$_7$, due to the low 
moment size, they scale as the square of the moment, and thus larger defective moments would tend to produce a strong, random perturbation on the dipole sum. 

\begin{figure}[h]
\centering
\includegraphics[width=0.9\columnwidth]{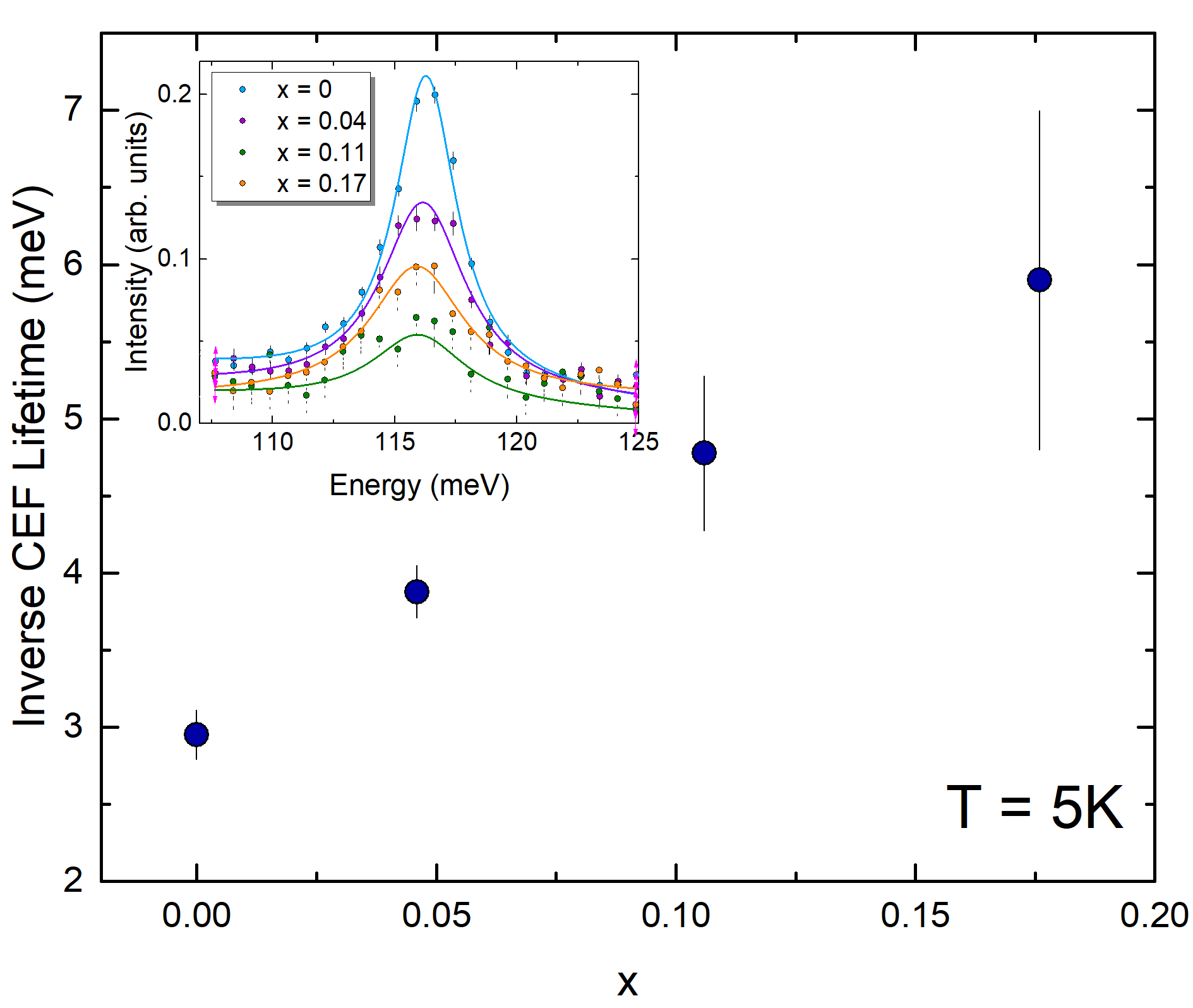}
\caption{\label{fig: 5}(color online) The main panel shows the systematic broadening of the CEF intrinsic energy width which is observed with increased ``stuffing", $x$, in $\rm Yb_{2+x}Ti_{2-x}O_{7-y}$, as obtained from the Lorentzian lineshape analysis discussed in the text. The inset shows the inelastic neutron scattering near the $\sim$ 116 meV CEF transition and the resulting fits performed with a Lorenztian lineshape.
}
\end{figure}

It is also clear that the \textit{A} site CEF transitions develop significant energy broadening with increasing stuffing. This can be broadly appreciated in Fig.~\ref{fig: 3}, and is examined quantitatively in Fig.~\ref{fig: 5}, where attention is focussed 
on the $\sim$ 116 meV \textit{A} site Yb$^{3+}$ CEF transition, which is well separated in energy from any other transition for all powder samples.  The energy width of the CEF excitations can be examined by fitting the data, 
shown in the inset of Fig.~\ref{fig: 5}, utilizing a damped harmonic oscillator (DHO) line shape for the 116 meV CEF transitions. At the energy transfers and temperatures of interest, the DHO can be approximated by a single Lorentzian, 
the form of which is given by,

\begin{equation}\label{eq: 1.6}
L(E) = \frac{1}{\pi}\frac{\frac{\Gamma_{obs}}{2}}{(E - \Delta E)^2 + (\frac{\Gamma_{obs}}{2})^2},
\end{equation}

This is a Lorentzian function of energy with width $\Gamma_{obs}$ centred on the energy of the CEF transition, $\Delta E$. This form convolutes both the intrinsic energy width, and that arising from the instrumental resolution, which are assumed to add in quadrature.  The intrinsic energy width or inverse lifetime of the 116 meV CEF excitation for each of the stuffed powder samples was extracted from this analysis and is plotted as a function of stuffing, $x$,  in Fig.~\ref{fig: 5}. 

Figure~\ref{fig: 5} clearly shows the CEF excitations at low temperatures in the stuffed powder samples to display much larger energy widths than that of the stoichiometric sample.  
The trend for low temperature CEF inverse lifetimes to systematically increase with stuffing, previously reported for the x = 0 and x = 0.05 powder samples \cite{Gaudet} is seen to extend to the largest stuffing 
level studied, x = 0.18. 

In conclusion, new time-of-flight neutron spectroscopy allows the possibility of detecting and distinguishing CEF excitations in complex real materials with relatively low levels of defective environments, 
and we have demonstrated this for the quantum spin ice candidate pyrochlore magnet Yb$_{2+x}$Ti$_{2-x}$O$_{7-y}$. Such detailed information is particularly important for the case of Yb$_2$Ti$_2$O$_7$, 
as its ground state displays unusually strong sensitivity to stoichiometry.  Our results specifically show Yb$^{3+}$ moments in stuffed and oxygen deficient environments display Ising anisotropy, rather than the 
XY local anisotropy displayed by the stoichiometric moments.  Such defective Yb$^{3+}$ moments are also considerably larger than their stoichiometric counterparts, and these, at a minimum, would tend to 
randomize dipolar interactions.  Both of these manifestations of stuffing can be important for ground state selection in real samples of Yb$_{2+x}$Ti$_{2-x}$O$_{7-y}$, and may underlie the ground state's 
extreme sensitivity to stoichiometry in this family of quantum magnets.

Research conducted at McMaster University was supported by the Natural Sciences and Engineering Research Council of Canada (NSERC). We acknowledge useful discussions with A. Aczel, L. Balents, G. Ehlers, M. D. Lumsden, S. E. Nagler and K. A. Ross. We are very grateful for the instrument and sample environment support provided during our inelastic neutron scattering measurements. The experiments which were performed at the Spallation Neutron Source at Oak Ridge National Laboratory was sponsored by the US Department of Energy, Office of the Basic Energy Sciences, Scientific User Facilities Division.

\bibliographystyle{apsrev4-1}
\bibliography{Yb2Ti2O7_Stuffed_Complete_Oct26}

\newpage
\onecolumngrid

\section{Supplemental Material: Crystal field excitations from $\mathrm{\mathbf{Yb^{3+}}}$ ions at defective sites in highly stuffed $\mathrm{\mathbf{Yb_2Ti_2O_7}}$}
        
In this supplemental information we show the details of the crystal field calculation with which we analysed the neutron scattering data. We also highlight several details of the inelastic neutron scattering spectrum as a function of temperature. Finally we discuss sample preparation and refinement of the neutron powder diffraction data for the two highly stuffed samples of $\rm Yb_{2+x}Ti_{2-x}O_{7-y}$.
        
\subsection{\label{sec: intro1} A. Crystal Field Program} 
        
In order to analyse the neutron scattering data and fit the Crystal Electric Field (CEF) excitations we developed a calculation based on the point charge model~\cite{Hutchings} and on the Stevens' formalism~\cite{Stevens}.  The former neglects the overlap between the orbitals and any relativistic corrections, while the latter is a mathematical tool to write an expansion of the Coulomb potential of the crystal based on the symmetries of the environment that surrounds the magnetic ion. In our samples, the magnetic rare earth ion is sitting at 3 different environments: stoichiometric $A$ sites, oxygen deficient $A^\prime$ sites and $B$ sites. Figure 2 in the main paper shows these 3 environments.  Notice that we rotated the reference system in order to align the local $\langle 111 \rangle$ direction along $\hat{z}$.

In general the Coulomb potential of the crystal can be expressed using a linear combination of tesseral harmonics as follows,

\begin{eqnarray}
V(x, y, z) & = \frac{q_j}{4 \pi \epsilon_{0}} \sum_{n=0}^\infty \frac{r^n}{R_{j}^{(n+1)}} \left[ \sum_{m} \frac{4 \pi}{(2n+1)} Z_{n m} (x_j , y_j , z_j) Z_{n m} (x, y, z) \right].
\end{eqnarray}

Here $q_j$ is the charge of the ligand, $R_j$ is the position of the ligand and $Z_{n m} (x_j , y_j , z_j)$ is the tesseral harmonic~\cite{Hutchings}.

If we centre our reference system on the magnetic ion, we can rewrite the previous equation in this way,
\begin{eqnarray}
V(x,y,z) = \frac{1}{4 \pi \epsilon_{0}} \sum_{n}^\infty \sum_{m} r^n \gamma_{n m} Z_{n m} (x, y, z),
\end{eqnarray}
where for $k$ ligands,
\begin{eqnarray}
\gamma_{n m} = \sum_{j=1}^{k} \frac{q_j}{R_{j}^{(n+1)}}\frac{4 \pi}{2n +1} Z_{n m} (x_j , y_j , z_j).
\label{eq: 1.3} 
\end{eqnarray}
  
Equation~\ref{eq: 1.3} gives the coefficients of the linear combination of the tesseral harmonics. For every point group, only a few terms in the expansion are non zero (see Ref.~\cite{Walter1984}), 
and these terms coincide with the number of Stevens Operators we use in our Hamiltonian.

The point group of both the scalenohedron and the trigonal anti-prism is $D_{3d}$ and thus, following Prather's convention~\cite{Prather}, only the terms $Z_{20}, Z_{40}, Z_{43}, Z_{60}, Z_{63}$ 
and $Z_{66}$ survive in our expansion. This convention states that the highest rotational $C_3$ axis of the system must be rotated along $\hat{z}$ and one of the $C_2$ axis along $\hat{y}$,
assuring in this way that we have the minimum number of terms in the Coulomb expansion.

Finally we can use the so called ``Stevens Operators Equivalence Method" to evaluate the matrix elements of the crystalline potential between coupled wave functions specified by one particular value
of the total angular momentum $J$. This method states that, if $f(x,y,z)$ is a Cartesian function of given degree, then to find the operator equivalent to such a term one replaces $x$, $y$, $z$ with $J_x$, $J_y$, $J_z$ respectively, keeping in mind the commutation rules between these operators. This is done by replacing products of $x$, $y$, $z$ by the appropriate combinations of $J_x$, $J_y$, $J_z$, divided by the total number of combinations. Note that, although it is conventional to use $J$ or $L$ in the equivalent operator method, all factors of $\hbar$ are dropped when evaluating the matrix elements.

As we are studying the ground state (GS) of a rare-earth system, without an external field applied, $S^2$, $L^2$, $J^2$ and $J_z$ are good quantum numbers. Thus the Crystal Field Hamiltonian can now be written as:
\begin{eqnarray}
H_{CEF} = const. \sum_{nm} \left[\frac{e^2}{4 \pi \epsilon_0} \gamma_{nm} \langle r^n \rangle \theta_n \right ] O_{n}^m 
= \sum_{nm} \underbrace{[A_{n}^m \langle r^n \rangle \theta_n  ]}_{B_{nm}} O_{n}^m = \sum_{nm} B_{nm} O_{n}^m,
\label{eq: 1.4}
\end{eqnarray}
where $\gamma_{nm}$ is the same coefficient as in Eq.~\ref{eq: 1.3}, $e$ is the electron charge, $\epsilon_0$ is the vacuum permittivity, $\langle r^n \rangle$ is the expectation value of the radial part of the 
wave function, $ \theta_n$ is a numerical factor that depends on the rare earth ion~\cite{Hutchings}, $const.$ is a constant to normalize the tesseral harmonics and $O_{n}^m$ are 
the Stevens Operators.

The terms $A_{n}^m \langle r^n \rangle \theta_n$ are commonly called Crystal Field Parameters, and they coincide with the parameters we fit in our calculation.
A general form of the Hamiltonian for our system is therefore:
\begin{eqnarray}
H_{CEF} = B_{20} O_{2}^0 + B_{40} O_{4}^0 +B_{43} O_{4}^3 +B_{60} O_{6}^0 +B_{63} O_{6}^3 +B_{66} O_{6}^6.
\label{eq: 1.5}
\end{eqnarray}      

It is easy to verify that the equations are not linear, so we cannot write a closed system to solve the problem and identify a unique solution. In general a common way to solve non linear equations is to
create a function that closely approximates the result. We thus decided to use our Hamiltonian as a function of six CEF parameters, that are simultaneously fit to quantities of interests 
such as the energy of the excitations, the spectrum and the relative intensities of the levels. The quantity that the calculation minimizes is:
\begin{eqnarray}
\chi^2 = \sum_i \frac{(\Gamma_{obs}^i - \Gamma_{calc}^i)^2}{\Gamma_{calc}^i},
\label{eq: 1.7}
\end{eqnarray}
where $\Gamma_{calc}$ is the calculated quantity of interest and $\Gamma_{obs}$ is the observed quantity.

Following this spirit, the logic of the calculation is the following:
\begin{enumerate}
\item Starting with an initial set of CEF parameters, that can be calculated from first principles or taken from literature, we diagonalize our CEF Hamiltonian.
\item The eigenvalues are rescaled respect to the GS energy and we calculate and normalize the intensities and the CEF spectrum .
\item $\chi_{tot}^2 = \chi_{Energy}^2 + \chi_{Intensity}^2 + \chi_{Spectrum}^2  $ is calculated using Eq.~\ref{eq: 1.7}.
\item The procedure is iterated using another set of CEF parameters in order to minimize $\chi_{tot}^2$ until we converge on a solution which best estimates the experimental results.
\end{enumerate}

The minimization algorithm is robust and it assures the convergence towards a global minimum. The final CEF parameters are then used to calculate the spectrum for a direct comparison with the data set. Table~\ref{tab: 1} shows the best CEF parameters which were found minimize $\chi^2$ in the fitting procedure along with the energy eigenvalues corresponding to the CEF excitations of $\mathrm{Yb^{3+}}$ ions out of the ground state at the $A$, $A^{\prime}$ and $B$ sites.

\begin{table}[h!] 
\centering
\begin{tabular}{c|c|c}
\hline \hline
\multicolumn{3}{c}{Crystal Field Parameters (meV)} \\
\hline
\textit{A} Site & $A^{\prime}$ Site & \textit{B} Site \\
\hline
$\begin{array}{ccc} B^0_2 = 1.1 \\ B^0_4 = -0.0591 \\ B^3_4 = 0.3258 \\ B^0_6 = 0.00109 \\ B^3_6 = 0.0407 \\ B^6_6 = 0.00727 \end{array}$ &   
$\begin{array}{ccc} B^0_2 = -3.9860\\ B^0_4 = -0.002186\\ B^3_4 = 1.0655\\ B^0_6 = 0.001533\\ B^3_6 = 0.049192\\ B^6_6 = 0.01666\end{array} $ &
$\begin{array}{ccc} B^0_2 = -4.8744 \\ B^0_4 = -0.1407 \\ B^3_4 = 1.47542 \\ B^0_6 = -0.004862\\ B^3_6 = -0.1117 \\ B^6_6 = 0 \end{array} $ \\ \hline
\multicolumn{3}{c}{Calculated Spectrum (meV)} \\ \hline
$\begin{array}{ccc} 0.0~(d) \\ 76.72~(d) \\ 81.76~(d) \\ 116.15~(d) \end{array} $ &
$\begin{array}{ccc} 0.0~(d) \\ 90.17~(d) \\ 161.38~(d) \\ 179.36~(d) \end{array} $ &
$\begin{array}{ccc} 0.0~(d) \\ 130.98~(d) \\ 181.79~(d) \\ 358.14~(d) \end{array} $ \\ \hline \hline 
\end{tabular}
\caption{
\label{tab: 1}
Refinement of the CEF parameters and energy eigenvalues at each of the three Yb$^{3+}$ sites, from fits to the inelastic spectra data set at the three sites and relative energy levels.  All energy eigenvalues are doublets (d), as required by Kramers' theorem.}
\end{table}

\subsection{\label{sec: intro2}B. Temperature Evolution of the Crystal Field Excitations
        } 
        
\begin{figure}[h!]
\centering
\includegraphics[width=0.4\columnwidth]{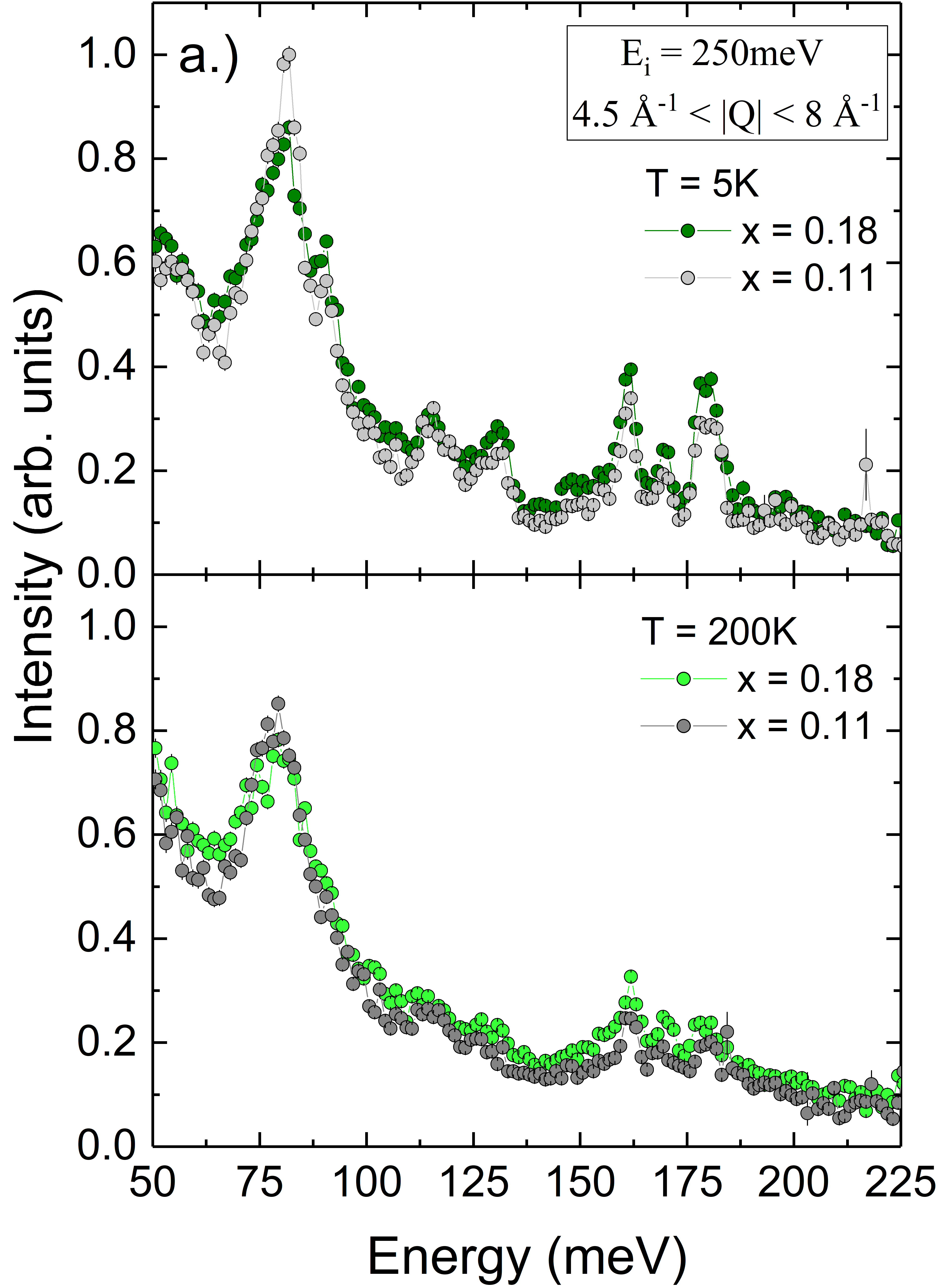}
\includegraphics[width=0.385\columnwidth]{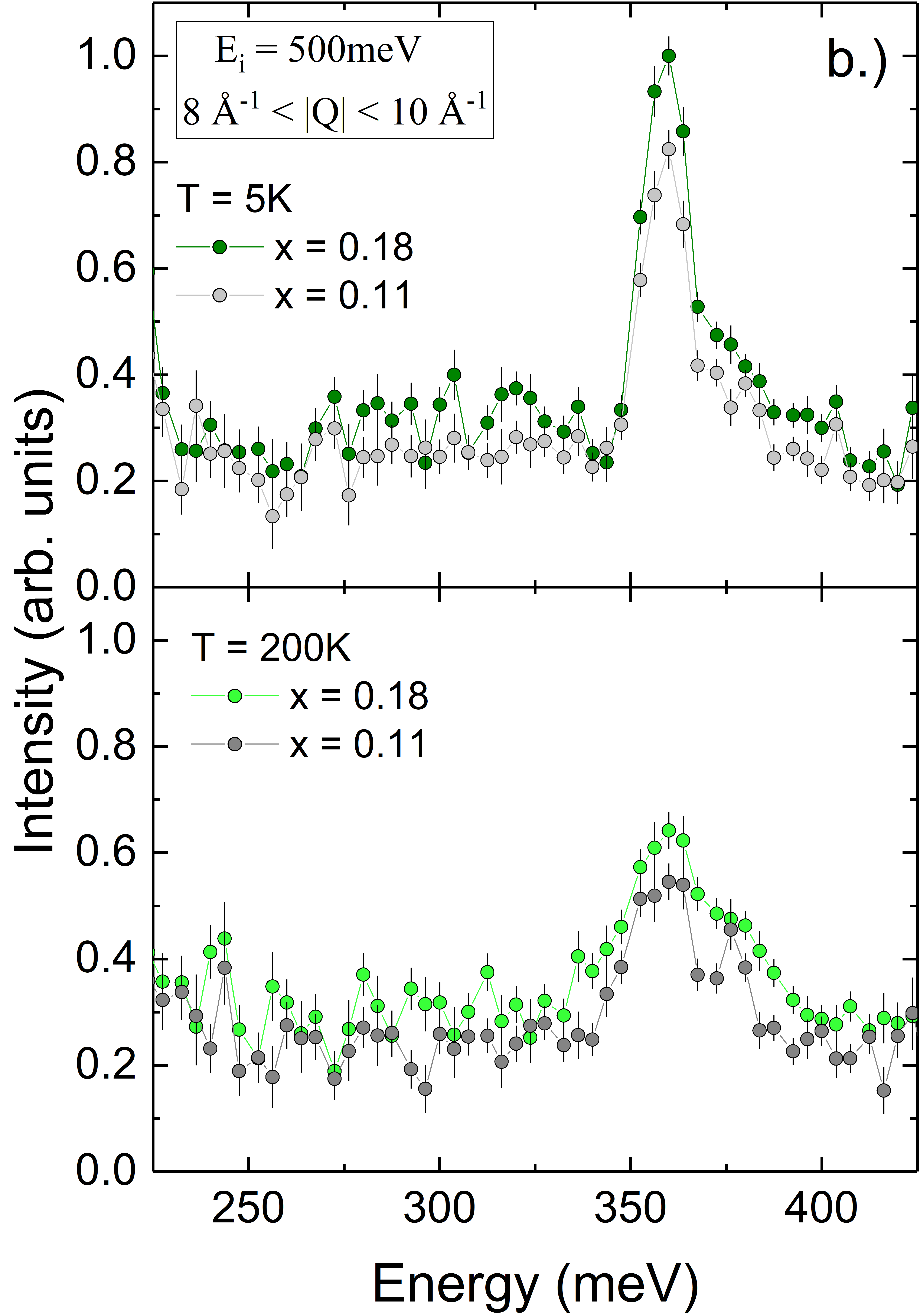}
\caption{\label{fig: 6}(color online) \emph{Temperature dependence of the inelastic neutron scattering from crystal field excitations: --} Comparison of the normalized intensities of the inelastic neutron spectrum at T = 5 K (top panels) and T = 200 K (bottom panels) for incident energies $E_i = 250$ and $E_i = 500$ meV. The feature at $\sim 170$ meV is not a CEF level since its intensity does not change with the temperature. By contrast the shape of the peak at $\sim 179$ meV  broadens at $200$ K, compared with $5$ K, consistent with the presence of two CEF levels (due to $A^\prime$ and $B$ transitions) close to each other in energy. The inelastic peak at $\sim 358$ meV arises due to Yb$^{3+}$ at $B$ sites.}
\end{figure}

Crystal field excitations have several important characteristics: as single ion properties the CEFs tend to be dispersion-less and the Q-dependence of their intensities is largely determined by the magnetic form factor of the magnetic ion involved.  They also display temperature dependence that reflects the population distribution of the CEF levels.  Given that the lowest energy CEF excited state is at $\sim$ 76 meV, and thus for all temperatures below room temperature, we expect no states above the ground state to be thermally populated.  
These features can be used to distinguish the real CEF levels from the background and from other elementary excitations, particularly phonons. In figure~\ref{fig: 6} we show the comparison of the CEF spectrum at T = 5 K (top panels) and T = 200 K (bottom panels) with an incident energy $E = 250$ and $500$ meV x = 0.11 and x = 0.18 samples.

As the temperature is increased the spectrum becomes broader in energy, in agreement with previous observations by Gaudet et al~\cite{Gaudet}. As mentioned above, this is not a thermal population effect but the result of the CEFs acquiring finite lifetimes, due to interactions with other excitations, notably phonons.  With the exception of the $A$ site Yb$^{3+}$ CEF excitations, the normalized intensity of the inelastic features in the spectrum are stronger for the x = 0.18 stuffed sample than for the x = 0.11 sample, as expected, reflecting the higher level of stuffing. 

The feature at $\sim 170$ meV is not a CEF level since its intensity does not change with the temperature. By contrast the shape of the peak at $\sim 179$ meV  becomes very broad 
at $200$ K consistent with the presence of two levels ($A^{\prime}$ and $B$ transitions) close one to each other.

The highest energy CEF feature at indicative of Yb$^{3+}$ at $\sim 358$ meV is due to stuffed $B$ site Yb$^{3+}$. This level has been previously predicted in Ref~\cite{Gaudet} and is now experimentally observed. The relatively high energy of this CEF is due to the fact that the oxygen ions surrounding the Yb$^{3+}$ at $B$ sites are closer to the Yb$^{3+}$ ions than is the case for either $A$ or $A^{\prime}$ site Yb$^{3+}$.
        
\newpage

\subsection{\label{sec: intro3}C. Sample preparation and characterization of the two highly stuffed Yb$_{2+x}$Ti$_{2-x}$O$_{7-y}$ samples
        } 

\subsubsection{1. Sample preparation}

Two single-crystals of Yb$_{2+x}$Ti$_{2-x}$O$_{7-y}$ with composition, $\mathrm{x = 0.11}$ and $\mathrm{x = 0.18}$ and dimensions $\mathrm{5~mm \times 6 mm \times 6~mm}$, were prepared by solid state reaction between pressed powders of $\mathrm{Yb_2O_3}$ and $\mathrm{TiO_2}$ which were sintered at 450 $^{\circ}$C for 15 hours with warming and cooling rates of 100 $^{\circ}$C/h. The sample preparation and characterization of the stoichiometric, x = 0, and x = 0.05 powders of Yb$_{2+x}$Ti$_{2-x}$O$_{7-y}$ are described elsewhere \cite{RossStuffing}. The purity of the starting powders of $\rm Yb_2O_3$ and $\rm TiO_2$ was close to $99.999\%$. To produce these highly stuffed, x = 0.11 and 0.18, samples of $\mathrm{Yb_2Ti_2O_7}$, a higher ratio of $\mathrm{Yb_2O_3}$ to $\mathrm{TiO_2}$ was used in comparison to what is conventionally used in order to produce stoichiometric samples of $\mathrm{Yb_2Ti_2O_7}$. The two single-crystals were grown at McMaster University by utilizing the floating zone image furnace technique, which is described elsewhere \cite{Dabko}. The growths were conducted in $\mathrm{O_2}$ gas with no overpressure and the growth rates were 7 mm/h and 8 mm/h for the x = 0.11 and x = 0.18 samples, respectively. The single-crystal samples were then pulverized using a Pulverisette 2 mortar grinder for 30 minutes each.

\subsubsection{2. Refinement of neutron powder diffraction data}

We discuss the results of our refinement for neutron powder diffraction data collected at POWGEN~\cite{POWGEN} at T = 300 K in table~\ref{tab: 2} for the x = 0.11 and x = 0.18 stuffed samples. This refinement was performed using the crystallographic refinement software JANA2006~\cite{JANA}. Note that the pure x = 0 and lightly stuffed x = 0.05 stuffed samples have previously been characterized by powder diffraction techniques in Ref.~\cite{RossStuffing}.

\begin{figure}[h!]
\centering
\includegraphics[width=0.5\columnwidth]{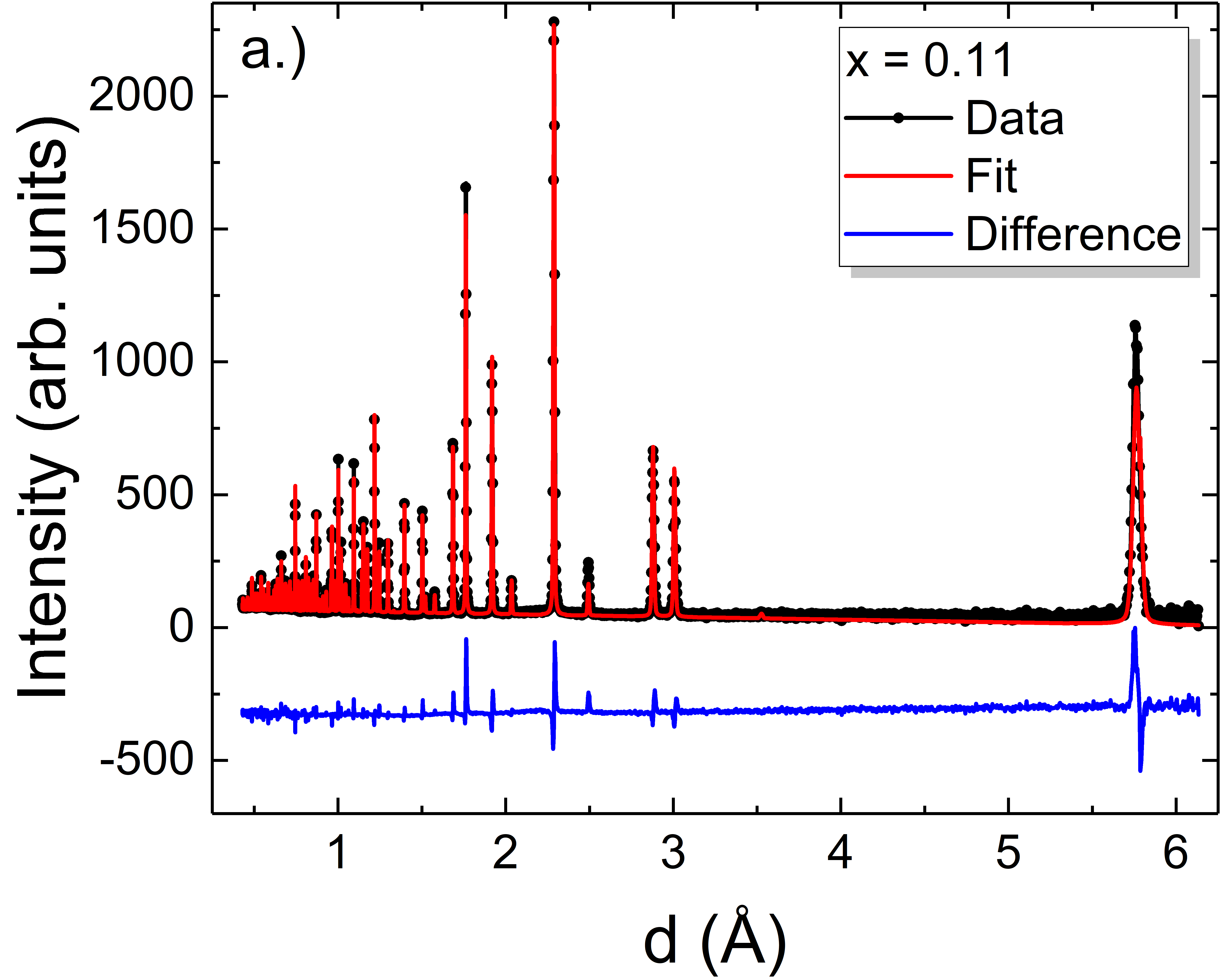}
\includegraphics[width=0.5\columnwidth]{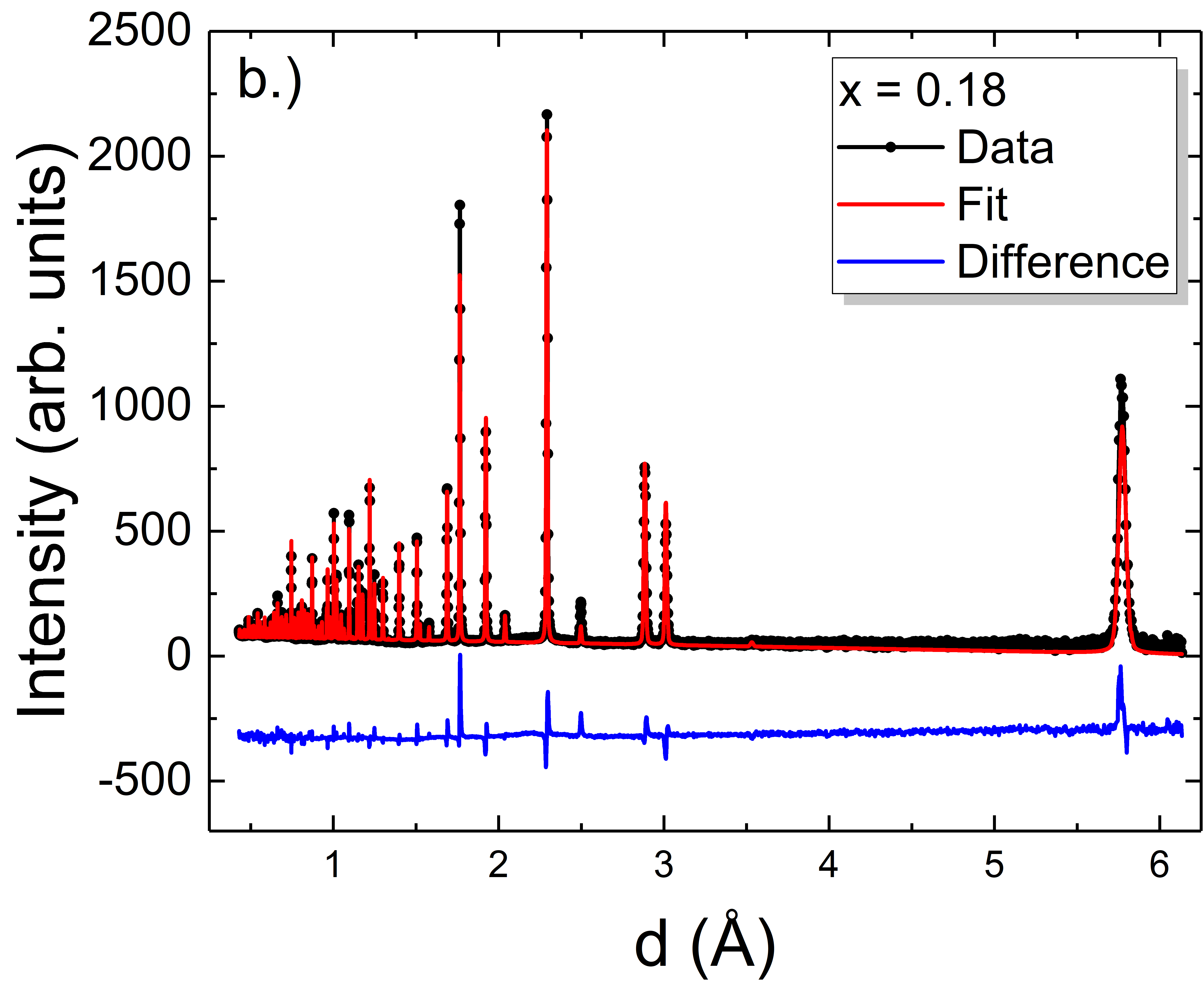}
\caption{\label{fig: 7}(color online) \emph{Rietveld refinement of the $\rm Yb_{2+x}Ti_{2-x}O_{7-y}$ stuffed crystals: -- } The Rietveld refinement has been performed assuming oxygen vacancies on the O(1) sites of the pyrochlore lattice, consistent with Ref.~\cite{Sala2014}. This model gives a good agreement with the experimental data with an $R = 5.1$ and $R = 5.2$ for the two crystals. The refined chemical formula for the two compounds are $\rm Yb_{2.106}Ti_{1.894}O_{6.952}$ and $\rm Yb_{2.176}Ti_{1.824}O_{6.883}$ giving a stuffing volume of x = 0.11 and x = 0.18 respectively, consistent with the approximate stoichiometry of the starting materials in the crystal growth. The unit cell parameter for the x = 0.11 stuffed sample is refined to be $a = 10.063(4)$ \AA, while that for the x = 0.18 is refined to be $a = 10.080(7)$ \AA.}
\end{figure}

The neutron diffraction powder spectra and best fits to the data are shown in figure~\ref{fig: 7}. The cell parameters arising from this refinement for the x = 0.11 stuffed sample is $a = 10.063(4)$ \AA, while for the x = 0.18, $a = 10.080(7)$ \AA. Table~\ref{tab: 3} highlights the systematic increase of the unit cell parameter, $a$, for all samples compared in this study. The fact that the length of the unit cell gets bigger as the stuffing, $x$, increases is a direct consequence of the oxygen vacancies; the Coulomb repulsion of the cations left unshielded by the vacancy tends to push all the ions away from each other increasing the size of the unit cell. Moreover our refinement showed that these vacancies are mainly located on the O(1) sites of the pyrochlore lattice, confirming the analysis in Ref.~\cite{Sala2014}. The refined chemical formula for the two compounds are $\rm Yb_{2.106}Ti_{1.894}O_{6.952}$ and $\rm Yb_{2.176}Ti_{1.824}O_{6.883}$ giving a stuffing of x = 0.11 and x = 0.18 respectively, in agreement with the approximate stoichiometry of the starting materials used in the crystal growth. 

\begin{table}[h] 
\centering
\begin{tabular}{c|ccc|c|c}
\hline \hline
\multicolumn{6}{c}{Neutron refinement x = 0.11 sample at 300 K} \\
\hline
Atom & x & y & z & Site & Occupancy \\
\hline
Yb & 0.625 & 0.625 & 0.625 & $16d$ & 1 \\ 
Ti & 0.125 & 0.125 & 0.125 & $16c$ & 0.947(2) \\ 
Yb & 0.125 & 0.125 & 0.125 & $16c$ & 0.053(2) \\ 
O(2) & 0.043(4) & 0.25 & 0.25 & $48f$ & 0.994(4) \\ 
O(1) & 0.5 & 0.5 & 0.5 & $8b$ & 0.988(3) \\ 
\hline \hline 
\end{tabular}
\qquad
\begin{tabular}{c|ccc|c|c}
\hline \hline
\multicolumn{6}{c}{Neutron refinement x = 0.18 sample at 300 K} \\
\hline
Atom & x & y & z & Site & Occupancy \\
\hline
Yb & 0.625 & 0.625 & 0.625 & $16d$ & 1 \\ 
Ti & 0.125 & 0.125 & 0.125 & $16c$ & 0.912(2) \\ 
Yb & 0.125 & 0.125 & 0.125 & $16c$ & 0.088(2) \\ 
O(2) & 0.458(3) & 0.25 & 0.25 & $48f$ & 0.984(10) \\ 
O(1) & 0.5 & 0.5 & 0.5 & $8b$ & 0.979(3) \\ 
\hline \hline 
\end{tabular}
\caption{
\label{tab: 2} This table summarizes Rietveld refinement results obtained from neutron powder diffraction experiments conducted at POWGEN \cite{POWGEN} on the two highly stuffed samples of $\mathrm{Yb_2(Ti_{2-x}Yb_x)O_{7-y}}$ with x = 0.11 and x = 0.18 at $\mathrm{T = 300 K}$.
}
\end{table}

\begin{table}[h] 
\centering
\begin{tabular}{c|c}
\hline \hline
\multicolumn{2}{c}{Refined lattice parameters} \\
\hline
Degree of stuffing, $x$ & a (\AA) \\
\hline
0.000(1) &  10.020(3) \\ 
0.046(4) & 10.029(4) \\ 
0.106(4) & 10.063(4) \\ 
0.176(8) & 10.080(7) \\
\hline \hline
\end{tabular}
\caption{
\label{tab: 3} The table shows results from Rietveld refinement for the degree of stuffing $x$ and the lattice parameter $a$ for the four compounds of $\mathrm{Yb_2(Ti_{2-x}Yb_x)O_{7-y}}$ studied. The values disclosed for the lattice parameters of the pure compound $\mathrm{x = 0.000(1)}$ and stuffed compound with $\mathrm{x = 0.046(4)}$ are the refined values at $\mathrm{T = 250 K}$ and were retrieved from Ref. \cite{RossStuffing}. The values of the lattice parameter obtained for the $\mathrm{x = 0.106(4)}$ and $\mathrm{x = 0.176(8)}$ were those obtained for the $\mathrm{T = 300 K}$ neutron diffraction data sets disclosed in table~\ref{tab: 2}.
}
\end{table}

Assuming that oxygen atoms are removed at random, we can perform a simple Monte Carlo (MC) simulation to calculate the relative preponderance of $A$ to $A^\prime$ sites in the lattice as a function of the stuffing level, $x$. Assuming that each $A$ and $A^\prime$ site contributes independently to the intensity of the spectrum, we can argue that the intensity of the transition at $90$ meV should be proportional to this ratio.

\begin{figure}[!htb]
\centering
\includegraphics[width=0.6\columnwidth]{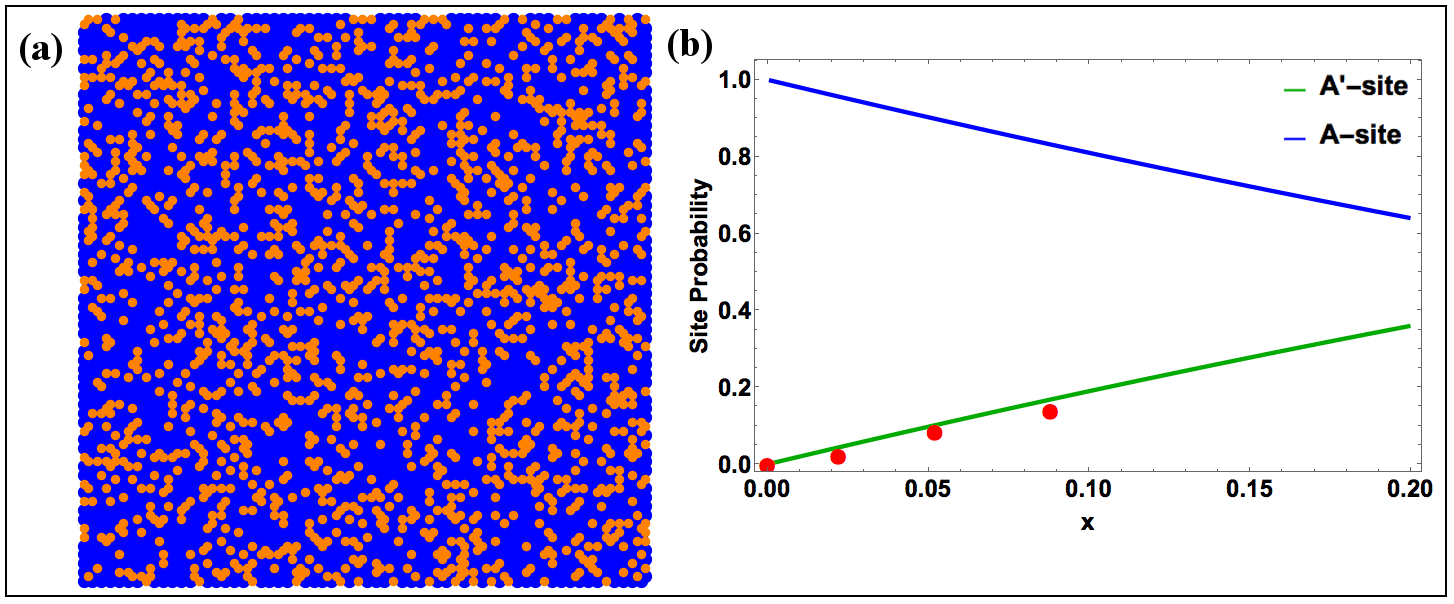}
\caption{\label{fig: 8}(color online) \emph{Preponderance of $A$ and $A^\prime$ sites within the pyrochlore lattice as a function of stuffing: -- } \textbf{(a)} Projection of the $64\times64\times64$ supercell used in the Monte Carlo simulation is shown.  The orange dots represent oxygen ions removed from the calculation. \textbf{(b)} Histogram showing the distribution of $A$ (blue line) and $A^\prime$ sites (green line) in the lattice as a function of the stuffing. The red points represents the experimental intensities of the $90$ meV CEF level extrapolated using the pure compound as background. This agreement confirms that this $90$ meV CEF transition originates from an $A^\prime$ site.}
\end{figure}

For this calculation we created a supercell consisting of $64\times64\times64$ unit cells filled with random vacancies located only at the O(1) position. Due to the symmetry of the pyrochlore lattice each Yb$^{3+}$ ion
at an $A$ site has only two O(1) ions as first nearest neighbour, thus we calculated how many ions have no vacancies and how many are affected by the stuffing. The calculation was repeated for $10000$ realizations of disorder. We show in figure~\ref{fig: 8} the results of this analysis, with the conclusion that the transition at $90$ meV originates from $A^\prime$ sites, and its intensity is directly proportional to the number of vacancies in the system.

\end{document}